\documentstyle[epsfig,astrobib]{mn-ab}

\title[Galaxy clustering in the Herschel deep field]
  {Galaxy clustering in the Herschel deep field}

\author[McCracken et.\ al]
{H.~J.~McCracken$^{1,3}$,T. Shanks$^1$, N. Metcalfe$^1$, R. Fong$^1$ 
\newauthor A. Campos$^{1,2}$\\
$^1$Department of Physics, University of Durham Science Laboratories, South Rd, Durham DH1 3LE.\\
$^2$ Instituto de Matematicas y Fisica Fundamental (CSIC)
Serrano 113 bis,  E-28006 Madrid, Spain \\
$^3$ Present address: Laboratoire d'Astrophysique de Marseille, 13376 Marseille Cedex 12, France\\}

\begin{document}
\def\wig{$\omega(\theta)$ }
\def\K{$K-$ band\ }
\def\bigwig{$\xi(r)$\ }
\def\sizex{16.0 cm}
\def\smallxsize2{7.0 cm}
\def\smallxsize3{5.0 cm}
\def\smallxsize{10.0 cm}
\def\smallysize{12.0 cm}
\maketitle

\begin{abstract}
  We present a study of the angular correlation function as measured in
  the William Herschel Deep Field, a high galactic latitude field which
  has been the subject of an extensive observing campaign from optical
  to infrared wavelengths. It covers $50$ arcmin$^2$ and with it we are
  able to investigate the scaling of the angular correlation function
  to $B\sim28$, $R, I\sim 26$,$K\sim20$ and $H\sim22.5$. We compare our
  measurements to results obtained from the smaller Hubble Deep Field.
  To interpret our results, we use a model which correctly predicts
  colours, number counts and redshift distributions for the faint
  galaxy population. We find that at fixed separation the amplitude of
  $\omega(\theta)$ measured in $BRI$ bandpasses is lower than the
  predictions of a model containing with no luminosity evolution and
  stable clustering growth in proper co-ordinates.  However, in the
  near-infrared bandpasses, our measurements are consistent with the
  predictions of an essentially non-evolving $K-$ selected galaxy
  redshift distribution.  In the range $B\sim27-28$ we find that our
  correlation amplitudes are independent of magnitude, which is
  consistent with the observed flattening of the number count slope and
  correspondingly slower increase of the cosmological volume element
  expected at these magnitudes.
  
  If our luminosity evolution models provide a correct description of
  the underlying redshift distributions (and comparisons to available
  observations at brighter magnitudes suggest they do), then our
  measurements in all bandpasses are consistent with a rapid growth of
  galaxy clustering ($0 < \epsilon < 2 $ in the normal parametrisation)
  on the sub-Mpc scales which our survey probes. We demonstrate that
  this rapid growth of clustering is consistent with the predictions of
  biased models of galaxy formation, which indicate that a rapid rate
  of clustering growth is expected for the intrinsically faint galaxies
  which dominate our survey.

\end{abstract}

\section{Introduction}
\label{sec:Introduction}

The projected two-point galaxy correlation function $\omega(\theta)$
has proved to be one of the most enduring statistics in observational
cosmology. This is a consequence of the relative ease with which it may
be measured; for each galaxy, all one requires is positions and
magnitudes.  Starting with the early studies of clustering in the local
universe using Schmidt plates \cite{1977ApJ...217..385G} to more recent
works using CCD-based detectors \cite{EBK,1992ApJ...399L..35P} these
studies have probed galaxy clustering to very faint magnitudes.
Normally, these surveys measure how the amplitude of the projected
angular correlation function at a fixed angular separation, $A_\omega$,
varies as a function of sample limiting magnitude -- the ``scaling
relation''. Usually, this relation has been parametrised in terms of
``epsilon models'' in which the three-dimensional correlation length
$r_0(z)$ scales monotonically with redshift
\cite{1977ApJ...217..385G,1978MNRAS.182..673P}. These models also
require a choice of cosmology and knowledge of the underlying redshift
distributions for each magnitude-limited sample.

In this paper we will investigate the projected angular clustering of
the faint field galaxy population. We characterise galaxy clustering as
a function of sample limiting magnitude in $BRIKH$ bandpasses. Our
primary dataset is a deep, ground based survey of an area called 'the
William Herschel deep field' (WHDF).  This has been described in
several recent papers \cite{MSCFG,2000MNRAS.311..707M}. Covering $\sim
50$~arcmin$^{2}$ this survey comprises an area $\sim 10$ times larger
area than the separate HDF-N and HDF-S fields. For comparison, we also
present a complementary analysis of clustering amplitudes measured in
these smaller fields, utilising the catalogues produced in Metcalfe et
al (2000). Although similar studies of $\omega(\theta)$ exist in the
literature \cite{EBK,RSMF,BSM,1996ApJ...469..519H,1997ApJ...490...11W}
our survey differs primarily in its depth ($B\sim28$) and broad
wavelength coverage (in this analysis we consider samples selected in
$BRIK$ bandpasses).

To interpret our results we use redshift distributions derived from the
luminosity evolution models we have described in our previous
papers (\cite{2000MNRAS.311..707M,MSCFG}.  These models are able to
reproduce all the observable quantities of the faint field galaxy
population (counts, colours, redshift distributions), at least for low
$\Omega_0$ universes and within current observational uncertainties
\cite{MSCFG}; it is these successes which give us confidence in using
our models as probe of the clustering history of the Universe, rather
than using our measurements of $A_\omega$ as a probe of the underlying
redshift distributions. In our models, high $\Omega_0$ Universes can be
accommodated by the model if we add an extra population of low
luminosity galaxies with constant star-formation rates which boost the
counts at faint ($B>25^m$) magnitude levels \cite{1997ApJ...488..606C}.
We also consider flat cosmologies with $\Lambda \neq 0$. For reference,
the scaling relation computed for a model with stable clustering and no
luminosity evolution is also presented.

Models such as those presented in this paper are relatively successful
in describing clustering measurements performed on deep blank-field
surveys like the one detailed in this work \cite{RSMF,BSM}. However,
observations of the clustering properties of Lyman-break galaxies
\cite{1996MNRAS.283.1388M} indicate that these objects have comparable
clustering properties \cite{1998ApJ...503..543G} to some classes of
locally observed galaxies, making such objects initially difficult to
understand in terms of this monotonic scaling of $r_0$ with redshift.
We will explain how these observations can be understood in the context
of the results presented in this paper.

Our paper is organised as follows: in
Section~\ref{sec:Herschel-Deep-Field} we describe in outline the
preparation of our datasets; in Section~\ref{sec:MethTech} we describe
the techniques we use to measure and analyse our data; in Section
\ref{sec:Meas-Ampl-Scal} we present our measurements of the projected
correlation function in five bandpasses in comparison with previous
work and investigate if our errors estimates are realistic; in
Section~\ref{sec:Interpr-Disc} we compare our correlation measurements
with the predictions of our evolutionary models; and finally, in
Section~\ref{sec:Conclusions-Summary} we outline the main conclusions
from this work.

\section{Observations and catalogues}
\label{sec:Herschel-Deep-Field}
Full details of the optical observations comprising the WHDF will be
presented in a forthcoming paper (Metcalfe et al 2000).
A subset of our infrared observations of the WHDF is described in
\citeN{2000MNRAS.311..707M} which comprises the $K<20$ UKIRT
observations.  Additional infrared observations at Calar Alto
Observatory produced a second catalogue limited at $H<22.5$ which will
be fully described in a separate paper (McCracken et al, in
preparation).  In this section we will briefly describe our object
detection and photometry techniques which are very similar to that used
in our previous galaxy counts papers
\cite{1991MNRAS.249..498M,1991MNRAS.249..481J,MSFR,2000MNRAS.311..707M}.
All our optical data discussed in this paper was taken at the William
Herschel Telescope (WHT), with the exception of a short $I-$band
exposure made at the Isaac Newton Telescope (INT).

After bias subtraction and flat-fielding, the sky background is removed
and isophotal image detection is carried out. These images are then
removed from the frame, replaced by a local sky value, and the
resulting frame smoothed heavily before being subtracted from the
original. This produces a very flat background. The isophotal detection
process is then repeated. A \citeN{1980ApJS...43..305K}-type
pseudo-total magnitude is then calculated for each image, using a local
value of sky.

Table \ref{tab:intro.limits} shows the magnitude limits for our fields.
As in our previous papers the minimum Kron radius is set to be that for
an unresolved image of high signal-to-noise, and the correction to
total is the light outside this minimum radius for such an image. Our
measurement limits give the total magnitudes of unresolved objects
which are a $3\sigma$ detection inside the minimum radius (which is
typically $\sim 1.4''$ for the WHDF data). Star-galaxy separation was
done on the $B$ frame using the difference between the total magnitude
and that inside a $1''$ aperture, as described in
\citeN{1991MNRAS.249..498M}. This enabled us to separate to
$B\sim24^m$. Some additional very red stars were identified from the
$R$ and $I$ frames. As the WHDF is at high galactic latitude the
stellar contamination should in any case be quite low. For the purposes
of measuring the correlation function, masks files were also
constructed to cover regions containing bright galaxies or stars.  The
area of the field affected by such bright objects is less than $10\%$
of the total.

\begin{table*}
\begin{center}
\begin{tabular}{lcccccc}
Filter& U & B & R & I & K & H\\
Limit (3$\sigma$) & 26.8 & 27.9 &26.3& 23.5/25.6&20.0 & 22.5\\
Area (arcmin$^2$) &  48.5 & 48.5 & 48.5 & 88.3/53.0  & 47.2 & 50\\
\end{tabular}
\end{center}
\caption{Photometric limits of the WHDF. The two magnitude limits in
  $I$ refer to two separate surveys, one carried out at the INT (and
  covering $88.3$ arcmin$^2$ and the other based on WHT data. }
\label{tab:intro.limits}
\end{table*}

Similar methods were also used to generate catalogues from the north
and south Hubble deep fields (i.e., we do not use any of the existing
HDF catalogues but use our own independently written object detection
and photometry software). One significant difference between the HDF
data and our ground-based data is of course their much higher
resolution. As described fully in Metcalfe et al (2000, in
preparation), we visually inspect all detections on our HDF N/S data in
an attempt to reduce the number of spurious entries in our catalogues.
We also carry out a 'reassembly' process in which multiple detections
on an individual galaxy are combined to produce a single detection.
This admittedly subjective procedure is unavoidable in the HDF
catalogues given the extremely high resolution of the data.

\section{Methods and techniques}
\label{sec:MethTech}
\subsection{Determining the angular correlation function}
\label{sec:Determ-Angul-Corr}

We use the normal estimator of \citeN{LS}, given in equation
\ref{eq:1.ls}.  Here we follow the usual notation where $DD$ indicates
the number of galaxy-galaxy pairs, $DR$ galaxy-random pairs and $RR$
random-random pairs for a given angular separation and bin width:

\begin{equation}
\omega ( \theta) ={\mbox{DD} - 2\mbox{DR} + \mbox{RR}\over \mbox{RR}}
\label{eq:1.ls}
\end{equation}

We find that, for a given survey sample, amplitudes measured by this
estimator are very similar to those computed with the
\citeN{1993ApJ...417...19H} estimator, $\omega(\theta)=(DDRR/DR^2)-1$.
The $DD/DR-1$ estimator, as used by \citeN{RSMF}, gives consistently
higher (at the $20-30\%$ level) values for $\omega(\theta)$, over all
our bins.  This has been found by other authors and is indicative of
the known biases inherent in this estimator
\cite{1993ApJ...417...19H,LS}.

For a range of magnitude-limited samples of each catalogue, \wig is
computed using equation \ref{eq:1.ls} for a series of bins spaced in
increments of $0.2$ in log($\theta$), where $\theta$ is in degrees. As
we have only observed one field we cannot use the field-to-field
variance to estimate the errors in each bin; instead we implement a
bootstrap-resampling technique
\cite{1984MNRAS.210P..19B,1986MNRAS.223P..21L}. In this method, the
error in each bin is computed from the variance of the estimator as
applied to a large ($\sim 200$) number of bootstrap-resampled
catalogues. As expected, these bootstrap errors are larger (normally
$\sim \times 2$) than the normal $\sqrt N$ Poisson counting errors.

To allow comparison with other workers, we fit our measured
correlations as a function of angular separation to an expression of
the form

\begin{equation}
\omega(\theta) = A_\omega (\theta^{-\delta} - C)
\label{eq:5}
\end{equation}

\noindent
where $A_\omega$ is the amplitude of \wig at $1^\circ$ and $C$ is the
``integral constraint'' term.  This term is a bias which arises
because we are using each catalogue to determine the mean galaxy
density and is particularly significant in our work because the area
which we survey is quite small. To calculate the integral
constraint we use the expression

\begin{equation}
C = {1 \over {\Omega^2}} \int \int \theta^{-\delta} d\Omega_1 d\Omega_2
\label{eq:6}
\end{equation}

\noindent
where $\theta$ is the angular separation of each galaxy pair and
$d\Omega_1$ and $d\Omega_2$ the solid angle subtended by each pair. If
we assume a power-law correlation function,
$\omega(\theta)\propto\theta^{-0.8}$ we may calculate this quantity for
our fields by direct integration. Typically we find $C\sim 13$
for the WHDF and $\sim40$ for the HDF (we must assume a slope for
power-law correlation function as we cannot calculate it directly from
this data; $-0.8$ allows us to compare our work with similar studies
in  the literature). 

\begin{figure} 
\centering
\centerline{\epsfxsize = \smallxsize
\epsfbox{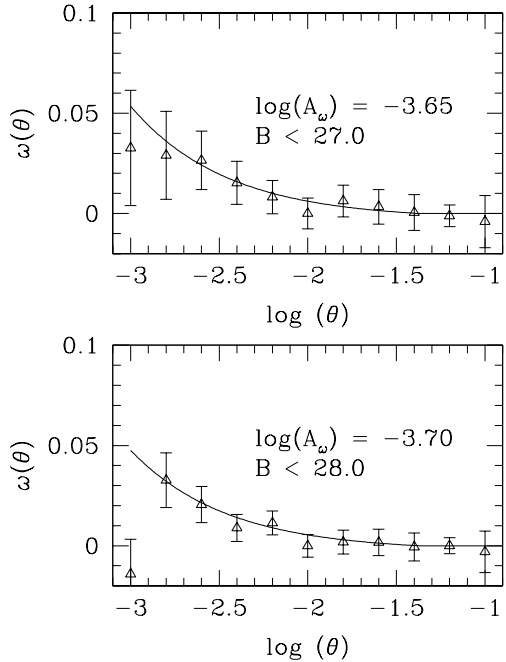}}
\caption{\wig as measured for samples limited at $B<27^m$ and $B<28^m$. The solid line shows the fit to
  $\omega(\theta) = A_\omega(\theta^{-0.8} - C)$ where $C$ is the
  ``integral constraint'' term described in the text and $A_\omega$ is
  the value of $\omega(\theta)$ at $1^\circ$}
\label{fig:linBcorr}
\end{figure}

The error on $A_\omega$, the overall fit, is determined from the method of
\citeN{marquard}, as described in \citeN{NR}. This method combines
errors on each bin in an independent manner to calculate the total
error of the fit.  Figure \ref{fig:linBcorr} shows fits made for the
$B-$band catalogue.

We determine correlation amplitudes for the Hubble deep field data
using a similar procedure. In this case we fit our final power law to
an average of the correlation function determined independently on each
of the three WFPC2 chips. For our NICMOS correlation amplitude, we
compute our correlation functions from the total numbers of pairs from
both surveys. For all these space-based data sets the field of view is
extremely small, and consequently the required integral constraint
correction is very large. Additionally, the small numbers of pairs
involved means that fits are generally dependent on three or fewer
bins, and for this reason our resulting correlation amplitudes
determined from these data should be regarded as upper limits on the
fitted amplitudes, rather than definitive measurements. In order to try
to reduce problems from ``merged'' objects as described in
Section~\ref{sec:Herschel-Deep-Field} we carry out our fits at angular
separation $>1''$.

\subsection{Modelling the correlation function}
\label{sec:Modell-Corr-Funct}

We would like to compare our measured correlation amplitudes with those
of model predictions. In order to do this we must assume a functional
form for the spatial correlation function. From the results of large
surveys
\cite{1977ApJ...217..385G,1983ApJ...267..465D,1990MNRAS.247P...1M} it
is found that $\xi(r)$ (the spatial correlation function) is well
approximated by $\xi(r)=(r_0/r)^{\gamma}$, at least for scales
$<20h^{-1}$~Mpc. Projecting a model for $\xi(r)$ onto the
two-dimensional distribution of galaxies measured by $\omega(\theta)$
involves integrating this function over redshift space using Limber's
formula \cite{1953Apj...117..134}.

We must parametrise the scaling of the correlation function with
redshift. Early papers \cite{1977ApJ...217..385G,1978MNRAS.182..673P}
assumed a scaling of the form 

\begin{equation}
 \xi(r,z)= h(z)\left({r_0\over r}\right)^\gamma 
\label{eq:scal}
\end{equation}

where 

\begin{equation}
h(z)=(1+z)^{-(3+\epsilon)}
\label{eq:hsca}
\end{equation}
(in this case $r$ is the proper distance); this relation has been used
in many recent observationally-motivated studies investigating the
projected two-point function
\cite{EBK,RSMF,BSM,1995ApJ...439..565I,1998ApJ...494L.137B}.

To derive an expression for $\omega(\theta)$, the projected correlation
function, we note that for small angles, the relation between
$\omega(\theta)$ and $\xi(r)$ becomes \cite{EBK}

\begin{equation}
\omega(\theta)= \sqrt \pi {\Gamma [(\gamma - 1) /2 ] \over
  {\Gamma(\gamma/2)}} {A \over {\theta ^{\gamma - 1} }} r_0^\gamma
\end{equation}
where $\Gamma$ is the incomplete gamma function, $\theta$ is the
angular separation and $A$ is given by
\begin{equation}
  A = \int^\infty_0 g(z)\left( dN \over {dz} \right)^2 dz / 
    \left[\int^\infty_0 \left( {dN \over {dz}} \right) dz \right]^2
\label{eq:aomega}
\end{equation}
where 
\begin{equation}
g(z)={h(z)\over{d_A^{\gamma-1}(z)\,(dr(z)/dz)}}
\label{eqnew}
\end{equation}
where $d_A(z)$ is the angular diameter distance and $dr(z)/dz$
is the derivative of the proper distance.

Analysis of the aforementioned large local redshift surveys suggests
that $\gamma = 1.8$, leading to three cases of interest to us:
clustering fixed in proper coordinates, in which case $\epsilon=0.0$;
clustering fixed in co-moving coordinates which gives $\epsilon=-1.2$.
Finally, the predictions of linear theory give $\epsilon=1.0$. This
formalism has been widely used in many papers which investigate the
clustering of faint field galaxies: see, for example, \citeN{PI3},
\citeN{1997ApJ...490...11W}.

As we have already noted, in these ``epsilon models'' characterised by
equation \ref{eq:scal} the co-moving galaxy correlation length
decreases monotonically with redshift (providing of course $\epsilon >
-1.2$, which produces models with clustering constant in co-moving
co-ordinates) However, several recent works have indicated that this
may not be a realistic assumption. In theoretical studies, both N-body
simulations \cite{1999ApJ...523...32C} and semi-analytic models
\cite{1999MNRAS.305L..21B,1999MNRAS.307..529K} indicate that the
co-moving galaxy correlation length decreases until $z\sim 1-2$ after
which it increases again. These theoretical studies
\cite{1998Natur.392..359G} also allow us to explain the high clustering
amplitudes observed for Lyman break galaxies at $z\sim3$
\cite{1998ApJ...503..543G,1998ApJ...505...18A} as a consequence of their
formation in highly biased environments. Furthermore, the clustering
growth is expected to be more rapid for less massive objects and and
for clustering amplitudes measured on smaller scales
\cite{1999MNRAS.305L..21B}.

Motivated by these works we also model our correlation amplitudes using
a modification of equation \ref{eq:scal}.  In place of the normal
epsilon parametrisation, we have used in the relativistic Limber's
equation a more general form for the evolution of $\xi(r,z)$, namely
\begin{equation}
\xi(r,z)= \left({r_0^{\rm com}(z)\over (1+z) r}\right)^\gamma
\label{eq:scalexi}
\end{equation}
where $r_0^{\rm com}(z)$ is the comoving correlation length at $z$.
Thus, we have used
\begin{equation}
h(z)=\left({r_0^{\rm com}(z)\over(1+z) r_0}\right)^\gamma
\label{eq:scale2}
\end{equation}
To illustrate the possible effect of modelling more exactly the
evolution of the correlation function, we have used the evolution seen
in the large N-body simulation of \citeN{1999ApJ...520..437K}; the
semi-analytic models mentioned above produce a similar form for the
evolution of $\xi(r,z)$ in their simulations. As our field sample is
dominated by spirals, we have therefore considered the haloes of the
simulation having velocity $v>120$km$^{-1}$. Also, as $\omega(\theta)$
for these deep fields has, as usual, been fitted to a $-0.8$ power law,
we have converted the Colin et al.  data to provide the same
correlation strength as a $-1.8$ power law for $\xi(r,z)$ at a comoving
separation of $0.3\,h^{-1}$~Mpc, which at the depths of our data here
corresponds roughly to the angular scale of our estimates for
$\omega(\theta)$.  Finally, to obtain the function, $r_0^{\rm com}(z)$,
a spline fit was made to the converted Colin et al.  data points with a
simple linear extrapolation to redshifts larger than the maximum
redshift, $z=5$, for which they have estimated the correlation function
for their simulation. Fig.~2 plots the resulting form of the evolution
used for $r_0^{\rm com}(z)$ normalised to $r_0$.  In using this in
Limber's equation, we have taken, as with Roche et al.  (1993),
$r_0=4.3\,h^{-1}$~Mpc, which is little different from the converted
Colin et al. value of $4.2\,h^{-1}$~Mpc.

\begin{figure}
\centering
\centerline{\epsfxsize = 10.0cm
\epsfbox{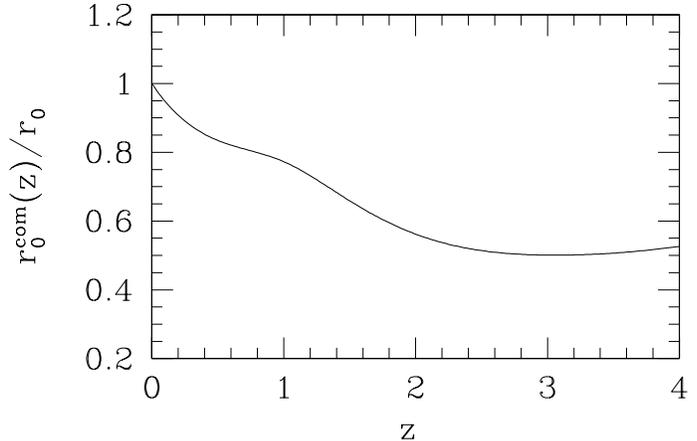}}
\caption{Normalised co-moving correlation length $r_0^{\rm com}(z)$
  as fitted to haloes of circular velocity $V>120$~km$^{-1}$ as
  identified by Colin et al in the large N-body simulation of Kravtsov
  and Klypin (1999). This line is closest to the predictions of
  $\epsilon=0$ model described in the text.}
\label{fig:rcomoving}
\end{figure}

\subsection{Calculating $dn/{dz}$}

From equation \ref{eq:aomega} we see that the amplitude of
$\omega(\theta)$ depends on the redshift distribution, $dn/{dz}$. To
produce these redshift distributions we employ a pure luminosity
evolution (PLE) model in which star-formation increases exponentially
with look-back time.  Earlier versions of these models are discussed in
our previous papers \cite{1991MNRAS.249..498M,MSFR,MSCFG}, and a full
description of the model used in this paper is given in
\cite{2000MNRAS.311..707M}.  In this paper we assume
$H_0=50$~kms$^{-1}$Mpc$^{-1}$, although changing the value of $H_0$
does not markedly affect our conclusions.  Two values of the
deceleration parameter $q_0 = 0.05$ and $q_0 = 0.5$, are adopted,
corresponding to open and flat cosmologies respectively.  The input
parameters to our models consist of observed {\it local\/} galaxy
parameters (namely, rest-frame colours and luminosity functions) for
each of the five morphological types (E/S0, Sab, Sbc, Scd and Sdm) we
consider in our models. These morphological types are divided into
early-type (E/S0/Sab) and spiral (Sbc/Scd/Sdm) and these two classes
are each given a separate star-formation history, parametrised in terms
of an e-folding time $\tau$. We compute the $k+e$ corrections using the
models of \citeN{BC}.  We could, in principle, sub-divide the spirals
into different morphological types each with different star formation
histories but for simplicity we do not; $(k+e)$ corrections for the
different types are fairly similar to each other in these models in any
case. Instead, taking a Sbc model as representative of all types we
produce the other types by normalising the Sbc track to the observed
rest-frame colours.  As in our earlier papers
\cite{1991MNRAS.249..481J,1991MNRAS.249..498M,MSFR,2000MNRAS.311..707M},
the normalisations of our luminosity functions are chosen to match the
galaxy counts at $B\sim18-20$ and we seek to explain the low number
counts at bright magnitudes from a combination of photometric errors
and anomalous galaxy clustering, rather than substantial and hence
unphysical evolution at low redshift in the luminosity of galaxies. Our
models also include the effects of the Lyman-$\alpha$ forest, and, for
spiral types, dust extinction corresponding to the Large Magellanic
Cloud as described in \citeN{Pei}. The model redshift distributions
produced are in good agreement with the redshift distributions of the
CFRS \cite{1995ApJ...455...50L} and from the Keck Hawaii redshift
survey \cite{CSH2}. To illustrate the effect which the inclusion of the
evolutionary corrections have on our computed correlation function
scaling relation, we also calculate an non-evolving redshift
distribution. This is produced by applying $k-$ corrections only to
each galaxy type.

\section{Measured Amplitudes}
\label{sec:Meas-Ampl-Scal}

\begin{figure*}
\centering
\centerline{\epsfxsize = \sizex
\epsfbox{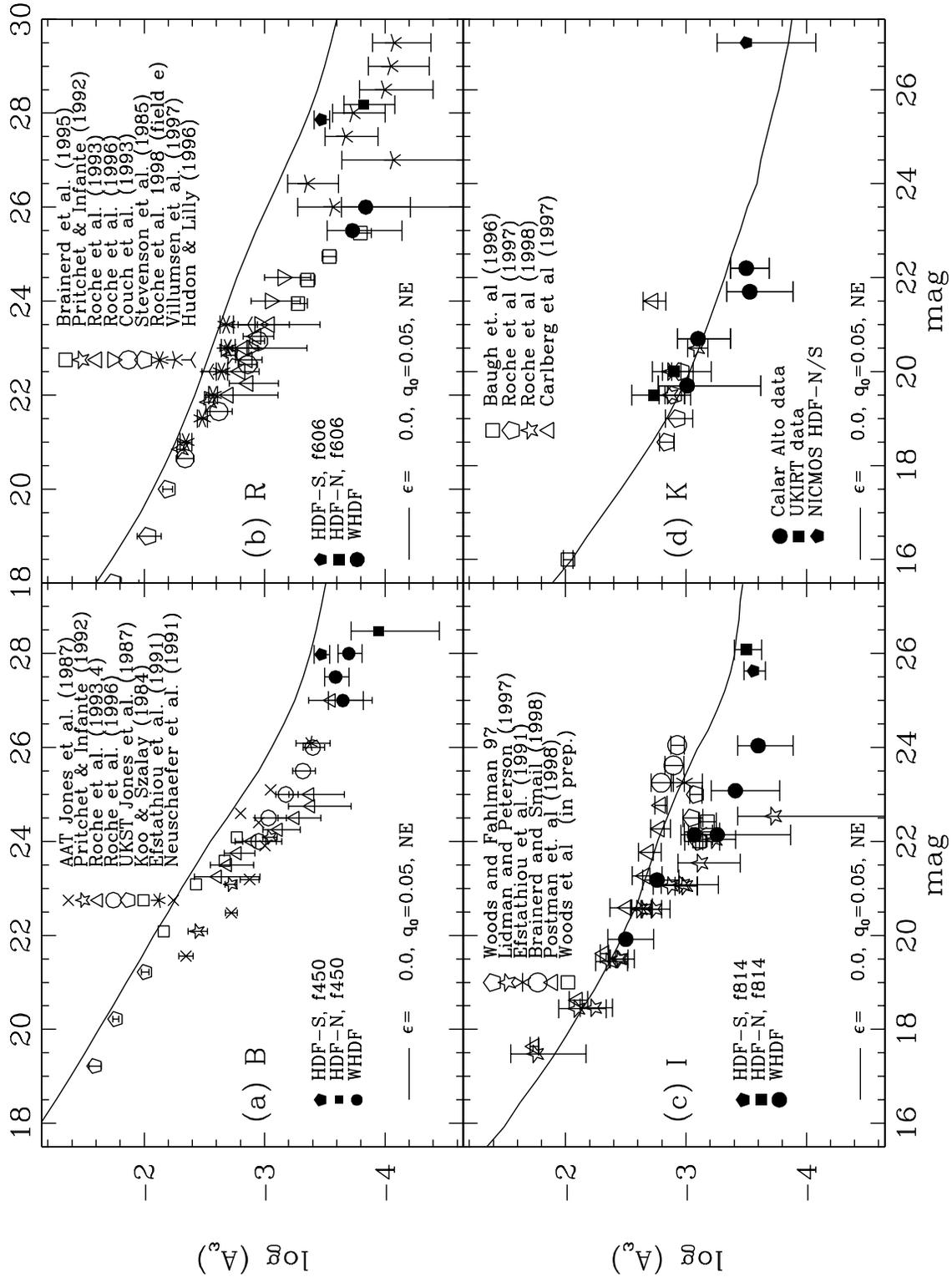}}
\caption{The logarithm of the amplitude of the angular correlation function
  $\omega(\theta)$ at one degree ($A_\omega$) in the WHDF (filled
  circles), HDF-N (filled squares) and HDF-S (filled pentagons) shown
  as a function of apparent magnitude for $BRIK$ selected samples
  (panels a--d). For $I$, correlations are plotted as a function of
  sample median magnitude. Open symbols show points from the
  literature. The solid line shows the predictions a non-evolving model
  with $\epsilon=0$ and with $r_0=4.3h^{-1}$~Mpc and $q_0=0.05$.  Error
  bars on our measurements are calculated by a bootstrap resampling
  technique, as described in Section~3.1~.}
\label{fig:allsca_ne}
\end{figure*}

In this Section we will present a comparison between our measurements
of $A_\omega$ and those in the literature. We defer an analysis of the
implications these measurements have for the growth of galaxy
clustering, as well as a discussion of our evolutionary models, to
Section~\ref{sec:Interpr-Disc}; here we present comparisons only with the
non-evolving, $\epsilon=0$, $q_0=0.05$ model.

In panels a--d of Figure \ref{fig:allsca_ne} we plot our fitted
correlation amplitudes extrapolated to one degree (filled symbols,
circles for WHDF and squares for HDF) as a function of sample limiting
magnitude for $BRIK$ bandpasses in comparison with measurements from
the literature (open symbols). The solid line shows the predictions of
the stable clustering, $\epsilon=0$ non-evolving (i.e., no luminosity
evolution) model , computed assuming $r_0=4.3h^{-1}$~Mpc and $q_0=0.05$
(This value of $r_0$ was chosen to produce the correct clustering
amplitude at brighter magnitudes as measured from early Schmidt plate
surveys \cite{1987Jones,1985MNRAS.213..953S}.  We adopt the same value
of $r_0$ for all bandpasses; in Section~\ref{sec:Growth-Clustering} we
discuss if this is an appropriate assumption for our data.)

\nocite{1996MNRAS.280..397R}  
\nocite{1991ApJ...382...32N}
\nocite{KS}

Starting with the $B-$ band, we note that here our WHDF sample reaches
extremely high galaxy surface density---approaching $\sim 10^{6}$
gal~$\deg^{-2}$ at $B=28^m$, and furthermore it probes to the highest
redshift; our low-$q_0$ evolutionary models indicate that by $B\sim 28$
we reaching $z_{med}\sim2$. Moreover, our measurements of the $B-$band
correlation function are significantly deeper than any previously
published work.  Our brightest bin, at $B<27.0$, is in agreement with
the correlation amplitude measured by \citeN{MSFR}. Faintwards of
$B=27$, our correlation amplitudes remain flat. The errors on our
fitted correlations in $B$ are relatively low in comparison with our
other bandpasses because at $B<28$ we detect $\sim 6000$ galaxies, more
than in any other bandpass. Our HDF-N/S clustering measurements are in
agreement with the measurement from the much larger area of the WHDF.

Our non-evolving models have some important differences with those used
in the earlier works of \citeN{RSMF} and \citeN{MSFR}.  Firstly, our
models include the effects of internal extinction by dust
(corresponding to $A_B=0.3$ mag, using the dust model of \citeN{Pei})
and reddening by the Lyman alpha forest (as modelled in
\citeN{1995ApJ...441...18M}). Both of these effects may become
significant at the very faintest magnitudes we reach, where $z_{med} >
2$.  Secondly, our $k-$ corrections are computed from the models of
\citeN{BC} for both our evolving and non-evolving models, whereas
\citeN{RSMF} and \citeN{MFS} used polynomial fits to the spectral
energy distributions of \citeN{P} for their non-evolving models. These
fits extend only to $z\sim2$ and are held constant at higher redshifts.
Thirdly, the redshift distributions in these earlier papers were
artificially truncated at $z=3$. The sum effect of these differences is
that in \citeN{RSMF} and \citeN{MSFR} the slope of the $A_\omega$ --
magnitude limit scaling relation remains constant whilst our slope
begins to decrease at $B\sim26$. By this magnitude limit the difference
between our predictions and these previous works is $\sim 0.2$ in
log($A_\omega$).

Our $R-$band correlations plotted in panel (b) of
Figure~\ref{fig:allsca_ne} reach $R<26$, although the number of
galaxies in this catalogue is much smaller ($\sim 300$) than in $B-$
and consequently our errors are larger. Our measured clustering
amplitude at $R<25.5$ agrees well with the faintest data point of
\citeN{BSM}; unfortunately, our survey area is too small to permit us
to check our clustering amplitudes with values from the literature
measured at brighter magnitudes such as the large, $\sim 2$~deg$^{2}$
CCD survey of \citeN{1999MNRAS.307..703R}. Our measured clustering
amplitudes in $R-$ in the WHDF are {\em much lower\/} than the
predictions of the non-evolving, stable clustering model. Our HDF
clustering measurements are in good agreement with the HDF clustering
measurements of \citeN{1997ApJ...481..578V}.

For our $I-$band measurements, shown in panel (c) of
Figure~\ref{fig:allsca_ne}, we follow the practice in the literature
and show correlation amplitudes as a function of sample {\em median\/}
magnitudes and not limiting magnitudes. We follow the same procedure
for our model correlation amplitudes which are plotted at the median
magnitude of each magnitude limited slice. In addition to our $I<26$
WHT data, we have a second, larger image taken at the INT which
overlaps the WHDF. This covers a total of $\sim 80$~arcmin$^{2}$ to
$I<23.5$ and allows us to determine $A_\omega$ from $I_{med}=20$ to
$I_{med}=22$ (the three brightest WHDF bins on the graph). The faintest
bin in this INT dataset is in agreement with our measurements from the
brightest bin of the WHT dataset. Furthermore, the preliminary result
from the large-area $0.2$~deg$^2$ survey of Woods et al. (in
preparation), shown as an open square, is agreement with our WHT
measurement. At $I_{med}\sim 26$, measurement from the HDF fields
appear to favour the lower values found in the WHDF. We note also that
fainter $I_{med}\sim 21$, our measurement are below the predictions of
the non-evolving $\epsilon=0$ model.

Faintwards of $I_{med}\sim 23$ a discrepancy emerges between our
measurements and two previously published studies. At $I_{med}\sim24$,
our WHDF clustering measurements are $\sim 5$ times lower than the
measurements made by \citeN{1998ApJ...494L.137B} over two slightly
smaller fields of area $\sim 30$~arcmin$^2$ at a similar limiting
magnitude.  At brighter magnitudes, our points are also below the
faintest bins of \citeN{1998ApJ...506...33P}.  This work is a
large-area CCD survey covering a contiguous $16$~deg$^{2}$ area and is
currently the most reliable determination of galaxy clustering over
wide angles and at intermediate $(z\sim1)$ depths. We defer a detailed
analysis of these differences until Section~\ref{sec:Errors-Biases}
where we will attempt to quantify if the discrepancies between our
survey and the works of \citeANP{1998ApJ...506...33P} and
\citeANP{1998ApJ...494L.137B} could be explained in terms of cosmic
variance effects.
  
Finally, we turn to an investigation of galaxy correlations for $K-$
selected samples. Until very recently measuring $\omega(\theta)$ at
near-infrared wavelengths was time-consuming and difficult as typical
detectors covered only $\sim 1$~arcmin$^2$. However, wide-format IR
arrays are becoming available making it now possible to conduct wide,
deep surveys of the near-infrared sky. The filled circles in panel (d)
of Figure~\ref{fig:allsca_ne} shows clustering amplitudes determined
from our faint, $H<22.5$, wide area ($\sim 50$~arcmin$^2$) Calar Alto
Survey are shown, which will be described fully in a forthcoming paper
(McCracken et al 2000, in preparation).  Similarly, also plotted are
clustering measurements from our $6'\times6'$ UKIRT IRCAM3 mosaic
\cite{2000MNRAS.311..707M}. At $K\sim27$ we have computed a single
point from NICMOS data taken as part of the north and south Hubble deep
fields program (we transform from $H$ to $K$ using a model
$(H-K)$ colour). We note that all our measurements are in agreement
with the predictions of our stable clustering, no luminosity evolution
model.

In plotting the $H-$ limited Calar Alto points on our $K-$ limited
scaling relation we make two assumptions: firstly, at $K\sim22$,
$(H-K)\sim0.3$; and secondly, for a given surface density, the
clustering properties of $H-$ selected and $K-$ selected galaxies is
identical. The first assumption seems reasonable, given that at $K\sim
20$, galaxies in our survey have $(H-K)\sim 0.3$ and it is unlikely
that they become significantly bluer by $K\sim 22$. The $K-$ selected
$(I-K)$ histograms shown in \citeN{2000MNRAS.311..707M} support this.
Also given that our Calar Alto $H<20$ $A_\omega$ agrees with our UKIRT
$K<19.5$ point, we conclude that our second assumption is also valid. 

Our points at $K=19-20$ agree with the survey of
\citeN{1998MNRAS.295..946R} and \citeN{1999MNRAS.307..703R}; however at
fainter magnitudes there is a discrepancy between our amplitudes and
the measurement of \citeN{1997ApJ...484..538C}. Once again, we defer a
detailed discussion of the possible explanation of these differences
until the following section. 

\subsection{Quantifying errors in the correlation function}
\label{sec:Errors-Biases}

\begin{figure*}
  \centering \centerline{\epsfxsize = \sizex
    \epsfbox{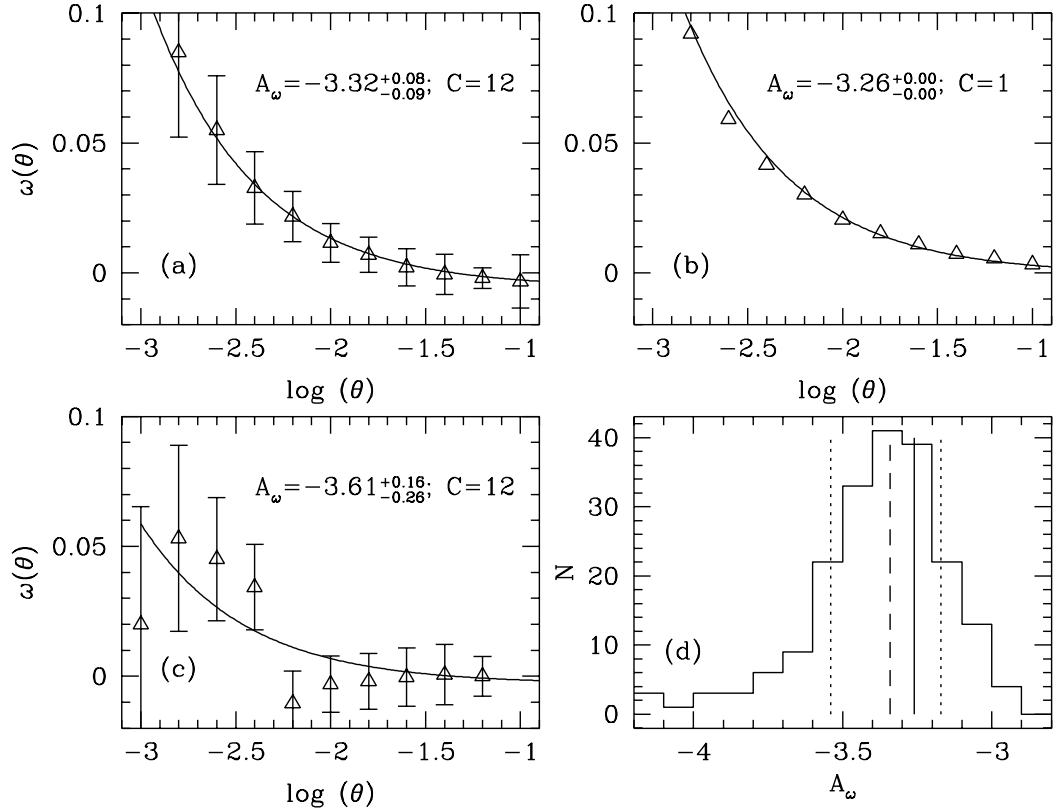}}
\vspace{-1.0in}
\caption{Results from simulations of a $6.25$~deg$^{2}$ area with
  the same surface density of objects as our $I<25$ catalogue. Panel
  (a) shows $\omega(\theta)$ measured from an average of two hundred
  sub-areas each covering $50$~arcmin$^2$ (corresponding to the size of
  the WHDF), with error bars calculated using the normal
  bootstrap-resampling technique; panel (b) shows $\omega(\theta)$
  determined from the full simulation. Panels (c) and (d) show
  $\omega(\theta)$ as measured at $I<25$ from the WHDF (again with
  error bars calculated using bootstrap-resampling) and the histogram of
  fitted values for $\log(A_\omega)$ from the simulations.  The dotted
  lines shown on the histogram represent $\pm 1\sigma$ deviations from
  the median value; the full simulation value is shown as the solid
  line and the average value as the dashed line. From this histogram we
  determine $\log(A_\omega)=-3.35^{+0.18}_{-0.17}$.}
\label{fig:wigerr}
\end{figure*}

In this Section we will investigate if we have estimated the magnitude
of our correlation function error bars correctly. The small size of our
field means our integral constraint (equation~(\ref{eq:6}) corrections
are large, and consequently accurate measurements of $\omega(\theta)$
are dependent on an accurate determination of this quantity. Our main
motivation is to see if we can explain the discrepancies between our
measurements of $A_\omega$ at $I<25$ and $K\sim21.5$ with those of
\citeN{1998ApJ...494L.137B} and \citeN{1997ApJ...484..538C}.  There are
already indications that such ``extra'' variance could be significant
at the depths of our survey.  \citeN{1998ApJ...506...33P} directly
address this question at shallower depths in their work which covers
$\sim 16$~deg$^2$. By extracting 250 independent
$16\arcmin\times16\arcmin$ fields from their survey (each of which is
five times larger than the WHDF but at a brighter limiting magnitude)
they find that the variance on $\omega(1\arcmin)$ is comparable to its
mean value of $\sim 0.045$, with extreme values reaching $\times 3$
this.  Furthermore, they suggest that as the error distribution for
$A_\omega$ is non-Gaussian, and skewed positively, there could be many
more areas in which $A_\omega$ is below the mean value, rather than
above it.

To quantify the amount of ``extra'' variance which could affect
clustering measurements in a very deep field like the WHDF we adopt a
simple approach and generate large mock catalogues using the method of
\citeN{1978AJ.....83..845S}. This is an purely empirical approach to
generate a hierarchically clustered distribution of points. We start
by placing within a sphere of radius $R$ a random distribution of
sub-spheres each of radius $R/\lambda$. Within each of these a further
$n$ spheres of radius $R\lambda ^2$ are added. This continues through $L$ 
levels; in our simulation we adopt $L=9$. The amplitude of the
correlation function is fixed by the number of centres used and the
fraction of the total number of points which are retained; these
quantities must be determined by trial and error. 

We measure the variance on the correlation function for many subsamples
of this catalogue. We start by generating a catalogue covering
$6.25$~deg${^2}$ with the same surface density of objects as in our
real catalogue at $I<25$ (corresponding to $\sim 7.5\times10^5$
galaxies). Next, we measure $\omega(\theta)$ over the full simulated
catalogue area.  Our aim is to produce a catalogue for which the fitted
correlation amplitude $\log (A_\omega)$ at $I<25$ is midway between the
result of \citeANP{1998ApJ...494L.137B} ($\log
(A_\omega)=-2.93^{+0.05}_{-0.06}$) and our own ($\log
(A_\omega)=-3.61^{+0.16}_{-0.26}$). Once a catalogue with the desired
correlation amplitude is produced it is randomly sub-sampled to produce
200 sub-areas each of which has the same field of view ($\sim
50$~arcmin${^2}$) and galaxy surface density at $I\sim25$ as the WHDF
(this translates to $\sim 2000$ objects per field). On each of these
sub-fields correlation amplitudes are measured using the same
parameters as for the real data set, and a histogram is computed using
each of these individual measurements of the simulated data.

Figure~\ref{fig:wigerr} presents results from one set of simulations.
Panel (b) in this Figure shows the measured correlation function for a
synthetic catalogue generated using the method outlined above (error
bars have not been plotted as they are smaller than the symbols for all
bins). For this catalogue we find $\log (A_\omega) =-3.26$ for an
integral constraint $C=1$. Panel (a) shows the average value of
$\omega(\theta)$ from two hundred sub-samples of this catalogue, as
well as the fitted value to this average which we find to be $\log
(A_\omega)=-3.32^{+0.08}_{-0.09}$. For comparison, the fit to our
$I<25$ observations is shown in panel (c); for the real data we find
$\log (A_\omega)=-3.61^{+0.16}_{-0.26}$.

This procedure tests several important aspects of our technique.
Because our simulated field is so large, the integral constraint
correction (Equation~\ref{eq:6}) which must be applied to it is much
smaller than the amount required for each of the individual sub-fields.
Given that the $A_\omega$ which we measure from the full survey agrees
to within the fitting errors to the $A_\omega$ determined from the
average of two-hundred sub-fields we conclude that errors arising from
an incorrect determination of the integral constraint are not
significant. However, the agreement between the sub-fields and
full-survey values is perhaps not surprising as both the catalogue and
$C$ were generated and calculated assuming a power-law slope $\delta$
for $\omega(\theta)$ of $-0.8$. There is some indication that $\delta$
becomes flatter at fainter magnitudes \cite{1998ApJ...506...33P}
although we are unable to test this with our current data set. A
flatter correlation function at fainter magnitudes would lead to an
underestimate of the integral constraint and a consequent 
  underestimate of $A_\omega$.

More significantly for this current work, however, is the large
dispersion we find for the fitted $A_\omega$'s from our simulated
fields.  Panel (d) of Figure~\ref{fig:wigerr} illustrates this. From
this diagram, $\log(A_\omega)=-3.35^{+0.18}_{-0.17}$ ($1\sigma$
errors). This corresponds to a linear error of $\pm 2.2\times10^{-4}$.
By comparison, our errors determined from our bootstrap resamplings are
$\pm 1.1\times 10^{-4}$; \citeANP{1998ApJ...494L.137B}'s errors are
$\pm 1.5\times 10^{-4}$.  Their points are an average of two widely
separated $\sim 30$~arcmin$^2$ fields and at $I<25$ contain
approximately the same numbers of galaxies as our catalogue. On the
basis of these simulations we conclude both our errors and those of
\citeANP{1998ApJ...494L.137B} underestimate the true error. Additional
simulations at higher amplitudes also have higher variances; a second
simulation at $\log(A_\omega)=- 3.1$ has a $1\sigma$ error of $\pm
2.6\pm10^{-4}$.  Adopting errors of this size, we find that our
correlation measurement is consistent with that of
\citeANP{1998ApJ...494L.137B} at the $2.5\sigma$ level. 


Turning to the $K$- selected correlation amplitudes plotted in
Figure~\ref{fig:allsca_ne} we note that at $K<21.5$ over $\sim
27$~arcmin$^2$ \citeN{1997ApJ...484..538C} measure $\log(A_\omega) =
-2.72^{+0.08}_{-0.10}$. This is also different from our work: at
$K<21.7$ we measure $\log(A_\omega) = -3.53^{+0.19}_{-0.36}$ in an area
of $\sim44$~arcmin${^2}$. The number density of galaxies at $K\sim21$
is approximately the same as at $I\sim25$, and the amplitude of
\citeANP{1997ApJ...484..538C}'s point is also within $\sim 40 \%$ of
\citeANP{1998ApJ...494L.137B}'s point. Moreover, as we have
described above, we find a larger variance on $\omega(\theta)$ for
simulations of higher amplitude. These considerations leads us to
conclude that the stated error bar on \citeANP{1997ApJ...484..538C}'s
measurement is also an underestimate of the true error. Furthermore,
we conclude that our ``low'' results at $I\sim25$ and $K\sim21$ are
\emph{not} inconsistent with the other results in the literature, given
the large error bars afflicting measurements of $A_\omega$ in fields of
this size and at these depths.

\section{Interpretation and discussion}
\label{sec:Interpr-Disc}
\begin{figure*}
\centering
\centerline{\epsfxsize = \sizex
\epsfbox{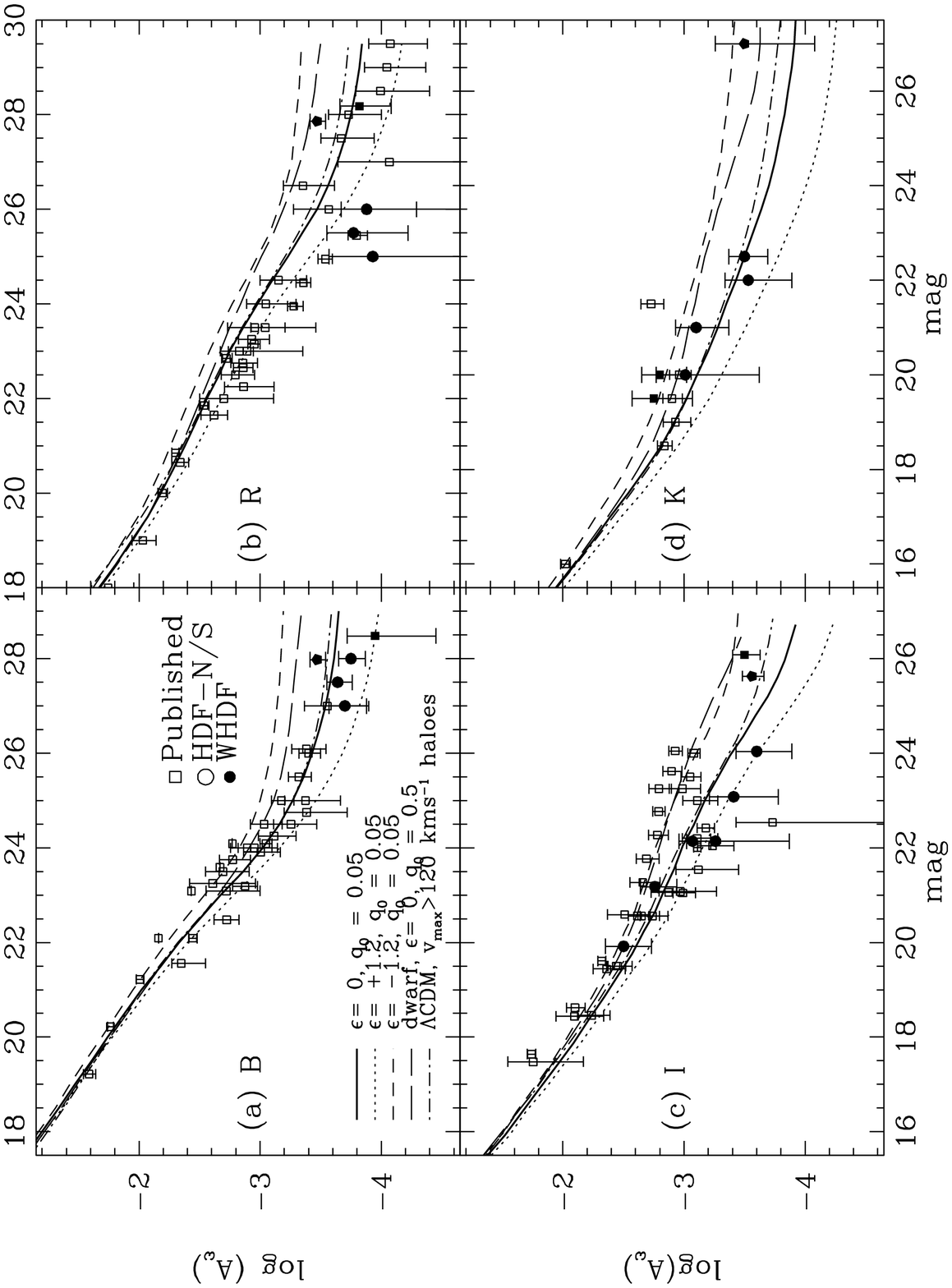}}
\caption{
  The logarithm of the amplitude of the angular correlation function
  $\omega(\theta)$ at one degree ($A_\omega$) in the WHDF (filled
  circles), HDF-N (filled squares) and HDF-S (filled pentagons) shown
  as a function of apparent magnitude for $BRIK$ selected samples
  (panels a--d). For $I$, correlations are plotted as a function of
  sample median magnitude; open symbols show points from the
  literature. Error bars on our measurements are calculated by a
  bootstrap resampling technique as described in Section 3.1~. Also
  shown are the predictions of our best-fitting evolutionary model for
  three values of the clustering growth parameter $\epsilon$ and for
  $r_0=4.3h^{-1}$~Mpc and $q_0=0.05$. The long dashed line shows the
  predictions of the $\epsilon=0$, $q_0=0.5$ dwarf-dominated model, and
  the dot-dashed line shows the predictions for the
  $\Omega_\Lambda$=0.7 case with dynamical evolution for haloes with
  rotation velocities $>$120 kms$^{-1}$. }
\label{fig:allsca_ev}
\end{figure*}
\nocite{1997ApJ...484..538C}
\nocite{1996MNRAS.283L..15B}

\subsection{Galaxy clustering models}
\label{sec:Fitting-data-with}

After a decade of study, the general characteristics of galaxy
evolution in the range $z=0-1$ have been broadly outlined, although
many specific details have yet to be worked out, such as parameter
dependence on morphology and intrinsic luminosity. Galaxy samples
selected in bluer bandpasses are dominated by starburst populations as
has been confirmed by many spectroscopic surveys
\cite{1991ApJ...369...79L,1995MNRAS.273..157G,CSH2}, whereas samples
selected in redder bandpasses show median redshifts and number count
distributions which are closer to the non-evolving predictions
\cite{MSCFG,CSH2}. This mirrors broad trends seen in studies of the
evolution of luminosity functions of colour-selected galaxy samples
\cite{1995ApJ...455..108L,1996MNRAS.280..235E,1997ApJ...475..494L,1999ApJ...518..533L},
although distinguishing between density and luminosity evolution in
these surveys is not straightforward. In interpreting our observations
in terms of luminosity evolution models we are well aware of the
limitations of these class of models, such as the difficulty in
correctly reproducing the properties of near-infrared selected samples.
Instead, we present our models as a simple parametrisation of the
observations using a model motivated by the star-formation behaviour of
the entire galaxy population. As shown in \citeN{MSF}, the integrated
star-formation history implied by our models agrees with current
estimates of the global star-formation history of the Universe.  The
agreement between our predicted redshift distributions and
observations at brighter magnitudes \cite{MSCFG,2000MNRAS.311..707M}
gives us confidence in using these models to investigate the growth of
galaxy clustering.

In panels a--d of Figure \ref{fig:allsca_ev} we plot our measurement of
$A_\omega$ for $BRIK$ bandpasses (filled symbols) in comparison with
the literature (open symbols), in addition to the predictions of our
evolutionary models with $\epsilon=-1.2,0,1.0$.  We also consider a
model with zero spatial curvature and an non-monotonic $r_0-z$
relation, as explained in Section~\ref{sec:Modell-Corr-Funct}. In
comparing these four graphs, it is interesting to notice how the
\emph{shape} of the scaling relation is qualitatively different from
bandpass to bandpass.  To understand the origin of these differences,
we start by emphasising that \emph{amplitude} of the correlation
function is directly related to the sample median redshift. From
Eq.~(\ref{eq:scal}) we see that $A_\omega$ depends on the width of the
redshift distribution, as well as its median value. Our model scaling
relations are therefore indicative of the median redshift of the model
population. Other workers have commented on this relation previously
\cite{KS}; in this section we will attempt to see if our two-population
luminosity evolution model can be used to explain the differences
between the observed scaling of the correlation function amplitudes.

Starting with the $B-$band scaling relation, we note that in this
bandpass in all magnitude slices, the model population is dominated by
spiral galaxies. Brighter than $B\sim22$, evolutionary effects are
negligible; however, faintwards of this, galaxies undergo $1-2$
magnitudes of brightening, and a significant high-redshift tail becomes
evident. For this reason the slope of the $B$-band number counts is
steepest in this range. The effect of this evolutionary brightening is
to cause the median redshift of the $B-$ selected redshift
distributions to increase rapidly faintwards of $B\sim24$; at
$B\sim22$, $z_{med}\sim0.3$, but by $B\sim24$, $z_{med}\sim0.7$ and by
$B\sim25$, $z_{med}\sim1$. This causes $A_\omega$ to drop rapidly below
the non-evolving prediction.  This extended $B$-band redshift
distribution, confirmed in the spectroscopic survey of \citeN{CSH2},
allows us to explain the observed clustering amplitudes without
recourse to positing a hypothetical, weakly clustered population
dominating the $B$- selected samples, as did some earlier authors
\cite{BSM,1995ApJ...439..565I}. In redder bandpasses, the situation is
slightly different; for example, in $R$ spiral evolution is more
gradual than in $B$-, causing a much less pronounced slope change in
the $A_\omega$-magnitude relation at $R\sim24$. Similar considerations
apply to the $I$-band. By the $K$-band, however, the form of the
$A_\omega$ limiting magnitude relation is determined primarily by the
early-type population; although the early-type counts turn over at
$K\sim20$ they still comprise more than half of the total galaxy
population faintwards of this.  Consequently, the $A_\omega$-limiting
magnitude relation has constant slope as these galaxy samples are
dominated by slowly-evolving early-type populations.

\subsection{Model comparison with observations}
\label{sec:comp-evol-models}

From a visual inspection of Figure~\ref{fig:allsca_ev}, our low-$q_0$
evolutionary model with $\epsilon=0-1$ provides the best fit to the
data at all magnitude limits; in the following Section we will present
a quantitative analysis of the growth of clustering implied by our
models.

At fainter magnitudes, certain models are disfavoured; for example, the
$q_0=0.5$, dwarf-dominated model, shown in Figure~\ref{fig:allsca_ev}
as the long dashed line, produces much higher correlation amplitudes
than the observations fainter than $B\sim26$. This is because, in
general, the median redshift of a magnitude-limited sample is lower for
a low-$q_0$ cosmology than for a high $q_0$ one, because the
differential volume element is smaller in the latter case.
Consequently, for the same value of $\epsilon$ and $r_0$, $A_\omega$ is
higher for $q_0=0.5$ than it is for $q_0=0.05$.  For this reason, our
$\epsilon=0$, $q_0=0.5$ dwarf model predicts higher clustering
amplitudes than our standard $\epsilon=0$ $q_0=0.05$ model. The exact
magnitude of the differences between the two models depends (in
addition to the cosmological considerations outlined above) on where
the median redshift of galaxies in the high-$q_0$ model is greater than
unity, where the star-formation rate for the dwarf types is constant,
or less than unity, and where they rapidly fade. This high clustering
amplitude leads us to reject the dwarf-dominated, $\epsilon=0$,
$q_0=0.5$ model.  Of course, this conclusion is dependent on the dwarf
population having the same intrinsic clustering properties as the
normal galaxy population, which may not be the case
\cite{RSMF,1995ApJ...439..565I}.

Earlier works conducted in $B-$ selected surveys suggest that by
$B\sim26$, $A_\omega$ ceases to decline and reaches a constant,
limiting value \cite{1996MNRAS.280..397R,MSFR,RSMF}; more recently,
\citeN{1998ApJ...494L.137B} claimed to have detected a similar
phenomenon at $I_{med}\sim 24$.  The scaling relation for our $B-$band
evolutionary model, shown in Figure~\ref{fig:allsca_ev}, flattens off
at very high number densities ($\log(N_{gal})\sim 6 \deg^{-2}$) and
faint magnitudes ($B\sim28$). At these limits, the relationship between
number density and median redshift levels off.  This is a consequence
of the steep faint-end slope luminosity function assumed for Scd and
Sdm spiral galaxies (which have \citeN{1976ApJ...203..297S} function
parameter $\alpha=-1.5$), which means that at fainter magnitudes one
observes intrinsically fainter rather than more distant galaxies, and
also of the reduction of the cosmological volume element at high
redshift. Our $B-$band correlations reach depths at which the
correlation function is expected to behave in this manner, and indeed
from $B=27.0$ to $B=28.0$ we do observe that the amplitude of
$A_\omega$ is almost independent of magnitude.

What are the implications of our $K-$ selected $A_\omega$
measurements? In \citeN{2000MNRAS.311..707M} we demonstrated how the
low median redshift found for $K-$ selected redshift distributions
\cite{CSH2} placed stringent limits on the amount of evolution
allowable in these bandpasses. In order for our PLE models to fit
\citeANP{CSH2}'s $K<19$ redshift distribution (which has a very low
median redshift, close to the predictions of a non-evolving model), we
had to assume a steep slope ($x=3$) for the initial mass function.
This reduces the amount of passive evolution at $K$- for early-types,
resulting in a total galaxy population with a lower median redshift
(Figure~5 of \citeN{2000MNRAS.311..707M}, illustrated how the variations
in IMF slope could affect the redshift distributions).

As we can see from Figure~\ref{fig:allsca_ev} our $\epsilon = 0$, low
$q_0$ evolutionary model incorporating this steep IMF slope fits the
observed clustering amplitudes for $K-$ selected samples quite well.
Therefore, the observed clustering amplitudes are consistent with the
underlying redshift distribution for $K$- selected samples which has a
low median redshift, close to the predictions of the non-evolving
model.  More significantly --- and beyond the spectroscopic limit of
the even the Keck telescope --- our Calar Alto data at $H\sim22$
indicates that even at these very faint magnitude levels, the $H$-
selected galaxy correlations are \emph{still} consistent with the
non-evolving prediction. Galaxy merging, however, could provide another
explanation for the low median redshift we infer for our $K$- sample. A
low median redshift for $K$- selected surveys is a general prediction
of the models of hierarchical galaxy formation
\cite{1998MNRAS.297L..23K}.


\subsection{Measuring the rate of clustering evolution}
\label{sec:meas-rate-clust}

In this Section we investigate what implications our measurements
$A_\omega$ have for the growth of galaxy clustering. As we have
commented earlier, the small angular size of the WHDF means we are
probing very small scales where the growth of galaxy clustering is
expected to be highly non-linear.  Most of the power in our correlation
function signal comes from our inner bins, at angular scales of $\sim
0.2'$; at $z\sim 1$, the typical median redshift of our samples, this
translates to linear dimensions of $\sim 0.05h^{-1}$~Mpc (for
$q_0=0.05$).  Additionally, how our samples are selected will affect
clustering amplitudes. In our flux-limited catalogues, a range of
galaxy luminosities will be present, and local redshift surveys have
shown that clustering amplitude may be a function of luminosity and
morphology \cite{1995ApJ...442..457L,1997MNRAS.285L...5T}. 

With these caveats in mind, in Table~5.1 we present the results for
best-fitting values for the parameters $r_0$ and $\epsilon$ in Equation
(\ref{eq:1.ls}) determined by $\chi^{2}$ minimisation using our WHDF
observations and the model ($\Lambda=0$, $q_0=0.05$) outlined in
Section~\ref{sec:Modell-Corr-Funct}. As before, we use redshift
distributions determined from our best-fitting evolutionary model.
Because of the strong co-variance between $r_0$ and $\epsilon$ it is
not possible to derive both parameters simultaneously from our dataset;
instead we investigate what values of $r_0$ and $\epsilon$ are implied
by ``reasonable'' choices of these parameters.

We wish to investigate what value of $\epsilon$ best fits our data and
to do this we fix $r_0$ to $4.3h^{-1}~$Mpc. This value of $r_0$ is
chosen to agree with angular correlation measurements determined from
large Schmidt plate surveys.  More recent work from local redshift
surveys approximately agrees with this value.  For example,
\citeN{1995ApJ...442..457L} find for the $b_J$ selected APM an $r_0$ of
$5.1 \pm0.2 h^{-1}$~Mpc and $\gamma=1.71$.  The $R-$ selected Las
Campanas Redshift Survey \cite{1997MNRAS.285L...5T} finds
$r_0=5.0\pm0.14h^{-1}$~Mpc.

\begin{table}
\begin{center}
\begin{tabular}{lccccc}
Bandpass&$\epsilon$&$r_0$ \vspace{2mm}\\
&($r_0=4.3$~$h^{-1}$~Mpc)&($\epsilon = 0)$\vspace{4mm}\\
$B$ & $0.40^{+0.35}_{-0.30}$ & $3.70^{+0.45}_{0.50}$\vspace{2mm}\\
$R$ & $2.65^{+1.30}_{-0.65}$ & $2.30^{+0.75}_{-1.05}$\vspace{2mm}\\
$I$ & $1.10^{+0.75}_{-0.50}$ & $3.35^{+0.45}_{-0.45}$\vspace{2mm}\\
$K$ & $0.05^{+0.65}_{-0.45}$ & $4.30^{+0.70}_{-0.80}$\vspace{2mm} \\
\end{tabular}
\caption{Best-fit values for $\epsilon$ for $r_0=4.3h^{-1}$~Mpc
  and for $r_0$ for $\epsilon = 0$, using redshift distributions
  computed from our best-fitting evolutionary model and assuming
  $q_0=0.05$. Errors quoted are $\pm 1\sigma$.}
\end{center}
\label{tab:fitse0r0}
\end{table}

In general, we find $\epsilon\sim0$ for $r_0=4.3h^{-1}$~Mpc and
$q_0=0.05$, from our own data alone. As we have already discussed, in
the $I$- band our points are different from those of
\citeANP{1998ApJ...494L.137B} at the $\sim 3\sigma$ level.  Their
survey subtends $\sim 30'$ on the sky, and reaches similar depths to
our own work, and so we would expect this survey to sample the same
environments as our own, and therefore to show broadly similar growth
of clustering. Combining \citeANP{1998ApJ...494L.137B}'s three $I$-
limited points with our own, we derive $\epsilon=0.70^{+0.70}_{-0.45}$,
again for $r_0=4.3h^{-1}$~Mpc.

\begin{figure}
\centering
\centerline{\epsfxsize = 10.0cm
\epsfbox{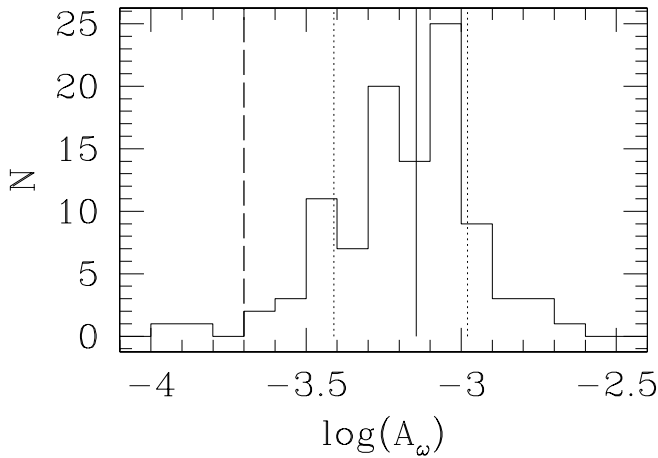}}
\caption{Histogram of fitted values of log($A_\omega$) carried out in 100
  sub-samples of a 1~deg$^{2}$ area containing $\sim 4\times 10^5$
  particles, representing the surface density of objects in our $B\sim
  27.5$ sample. The solid line shows the median value of the histogram,
  dot-dashed lines illustrate the $\pm 1\sigma$ confidence limits, and
  the dashed line shows the fitted value obtained from the WHDF.}
\label{fig:Berr}
\end{figure}

We have also carried out a simulation similar to those described in
Section~\ref{sec:Errors-Biases} to see how secure is our rejection of
the co-moving amplitude in the range $B=27-28$, the results of which
are presented in Figure~\ref{fig:Berr}. This simulation has the same
galaxy number density as our observations at $B\sim 27.5$ and covers an
area of $1$~deg$^2$. It contains a total of $\sim 4\times10^5$ galaxies
for an integral constraint $C=3.2$.  The simulation has
log$(A_\omega)=-3.12\pm 0.02$ (bootstrap errors), corresponding to the
amplitude of our co-moving evolutionary model at this magnitude. Errors
calculated by resampling 100 WHDF-sized fields over this area gives a
median $\log(A_\omega) = -3.15^{+0.16}_{-0.26}$ (1$\sigma$) and log$(A_\omega) =
-3.15^{+0.36}_{-0.46}$ ($2\sigma$). At $B\sim 27$ we measure
log$(A_\omega)=-3.7^{+0.13}_{-0.18}$ (bootstrap errors). Out of the 100
simulated fields, there are only two measurements at or below this
value, leading us to conclude that in this magnitude range our
measurement and the co-moving amplitude differs by at least $2\sigma$.

\subsection{The growth of clustering and biased galaxy formation}
\label{sec:Growth-Clustering}

How does this observed rate of clustering growth compare with
measurements from the literature? With deeper pencil-beam spectroscopic
surveys, it has become possible to measure $r_0$ at successively
earlier epochs and to use this to infer a value for $\epsilon$. Using a
statistically complete subsample of $591$ galaxies from the
Canada-France Redshift Survey \cite{1995ApJ...455...50L}, \citeN{FHLHT}
were able to measure the evolution of $r_0$ in the interval $0 \leq z
\leq 1$. They found $r_0(z=0.53)=1.5\pm0.09~h^{-1}~$Mpc (for
$q_0=0.05$), implying $0<\epsilon<2$.  \citeN{1997ApJ...484..538C},
using a sample of 248 galaxies, found that for $M_K\leq-23.5$ galaxies,
$r_0(z\sim0.6)=2.0^{+0.9}_{-0.2}~h^{-1}$~Mpc. At higher redshift,
\citeANP{1997ApJ...484..538C} derive $r_0(z\sim
0.97)=1.4^{+0.9}_{-0.2}~h^{-1}$~Mpc, which, combined with the lower
redshift points from their survey, leads to $\epsilon\sim0.2\pm0.5$.
The Canadian Network for Observational Cosmology (CNOC) have recently
completed a large field galaxy survey in the range $0\leq z \leq 0.7$
with a sample size of $\sim 10^4$ galaxies \cite{1998astro.ph..5131C}.
For luminous objects with corrected $R$-band absolute magnitudes of
$M_R^{k,e}< -20$ they find a slower clustering growth: $\epsilon =
-0.6\pm0.4$, with $r_0=5.15\pm0.15$, strongly excluding clustering
growth as rapid as $\epsilon \sim 1$.

The large size of the errors on our $\chi^2$ fit does not permit us to
make a detailed investigation of the dependence of $\epsilon$ with
sample selection. However, it is interesting that we find the rate of
clustering growth to be slowest for our $K$- selected survey, and that
our results are broadly consistent with those from the $K$- selected
sample of \citeN{1997ApJ...484..538C}. We expect our $K$- selected
galaxies to be good tracers of the underlying matter. $N$- body
simulations \cite{1997ApJ...490....1C} have shown that in a low-density
universe, the clustering of matter is expected to evolve as $\epsilon
\sim 0$.

Our finding that $0 > \epsilon > 2$, is in agreement with the
expectations from biased models of galaxy formation, which find that at
the $< 1 h^{-1}$~Mpc scales we are sensitive to, clustering growth is
relatively rapid (Figure 1. of \citeN{1999MNRAS.305L..21B};
\citeN{BBFC}). In comparison, at larger scales ($\sim 5h^{-1}$~Mpc) the
correlation function evolves much more slowly. At such separations, the
clustering pattern is ``frozen in'' as the galaxies are tracing
higher-mass haloes whose clustering evolution is close to
$\epsilon=-1.2$.  In Figure \ref{fig:allsca_ev} we see that that the
predictions from the non-zero lambda cosmology fitted to the growth of
clustering as observed in the simulation of \citeN{1999astro.ph..7337C}
is consistent with our observations and to the predictions of our
$\epsilon = 0 $ model. The rapid decrease of the co-moving correlation
length $r_0$ between $z=0$ and $z\sim2$ for small haloes of
$V>120$km$^{-1}$ is a prediction of biased models which find faster
clustering growth for intrinsically fainter galaxies. Even at brighter
than $B\sim26$ our samples are dominated by $\sim L^*$ galaxies
\cite{MSCFG}, unlike the Lyman-break galaxies of
\citeN{1996AJ....112..352S} which higher intrinsic luminosities than
this (The high resolution of \citeANP{1999astro.ph..7337C}'s simulation
and their adopted halo-finding algorithm makes it possible to locate
haloes within haloes and therefore to successfully match the halo
correlation function with the APM galaxy correlation function
\cite{1990MNRAS.242P..43M}. For this reason we assume an approximate
correspondence between these haloes and the intrinsically fainter
galaxies which dominate our samples at $B\sim28$.)

Figure~\ref{fig:allsca_ev} indicates that our observations are
consistent with models displaying the non-monotonic dip in the $r_0-z$
relation, and illustrates why the ``epsilon'' models have been so
successful in describing the observed scaling relation of
$\omega(\theta)$. If our models are correct, even to $B\sim28$ the
number of higher-redshift ($z>2.5$, more highly clustered galaxies
forms only a small fraction of total sample size (less than $<$5\%) and
this explains why scaling relations in which the galaxy correlation
length decreases monotonically $z\sim1$ can successfully match the
observations to $B\sim28$.

\section{Conclusions and Summary}
\label{sec:Conclusions-Summary}

In this paper we have presented a study of the projected two-point
angular correlation function $\omega(\theta)$ as measured in the WHDF
and compared our measurements to results obtained from the much smaller
North and South Hubble Deep Fields. The clustering amplitudes
determined from the HDF are consistent with those in the WHDF, but none
of our conclusions depend on our HDF measurements. In interpreting our
results, we have used redshift distributions from a model which
correctly predicts colours, number counts and redshift distributions
for the faint galaxy population.

We find that at a fixed separation the amplitude of $\omega(\theta)$
measured in $BRI$ bandpasses is {\em lower\/} than the predictions of a
low-$q_0$ model not containing luminosity evolution and in which
clustering growth is stable in proper co-ordinates. For our $K-$
selected samples, our correlation amplitudes are consistent with
predictions from models having a low median redshift; we also find
marginal evidence for a slower growth of clustering in these samples.

If our evolutionary models provide a correct description of the
underlying redshift distributions (and comparisons to available
observations at brighter magnitudes suggest they do), then our WHDF
clustering measurements are consistent with a clustering growth $0 >
\epsilon > 2$ on the small scales ($<1 h^{-1}$~Mpc) which we probe. We
have also shown that this result is consistent with prediction of
biased galaxy formation models which find faster clustering growth for
intrinsically fainter galaxies like those which dominate our deep
magnitude-limited surveys. We are able to use these rapid-growth
``epsilon'' models to successfully describe the clustering properties
of our samples because the highly-clustered high-redshift galaxy
population constitutes only a small fraction of the total galaxies
observed in our survey.

Finally, our constant correlation amplitude found at $B\sim27-28$ is
consistent with the expected reduction of cosmological volume element
at high redshift and a steeper faint end slope for spiral galaxies,
indicating that at these magnitude limits the median redshift of our
sample ceases to increase.

\section{Acknowledgements}
\label{sec:Acknowledgements}
H.J. McCracken wishes to thank Carlton Baugh, Ian Smail, Luigi Guzzo
and Olivier LeFevre for careful readings of an earlier version of this
manuscript.  NM acknowledges partial PPARC support.


\end{document}